\documentclass[12pt]{article}
\usepackage{amsmath,amssymb}
\usepackage{graphicx}% Include figure files
\usepackage{caption}
\usepackage{color}% Include colors for document elements
\usepackage{dcolumn}% Align table columns on decimal point
\usepackage{bm}% bold math
\usepackage{float}
\usepackage{hyperref} % hyperref must be loaded before apacite
\hypersetup{ colorlinks = true, citecolor = black, urlcolor = blue}

\definecolor{background-color}{gray}{0.98}

\usepackage{xspace}
\usepackage{relsize}
\usepackage{tikz}
\usepackage{booktabs}
\usepackage[utf8]{inputenc}
\usepackage[tight]{units}
\usepackage{textgreek}

\newcommand{\python}  {\texttt{python}\xspace}
\newcommand{\cpp}     {\texttt{C++}\xspace}

\newcommand{\adcc}      {\texttt{adcc}\xspace}
\newcommand{\adcman}    {\texttt{adcman}\xspace}
\newcommand{\pyscf}     {\texttt{pyscf}\xspace}
\newcommand{\psifour}   {\texttt{psi4}\xspace}
\newcommand{\psifournp} {\texttt{psi4numpy}\xspace}
\newcommand{\numpy}     {\texttt{numpy}\xspace}
\newcommand{\scipy}     {\texttt{scipy}\xspace}
\newcommand{\matplotlib}{\texttt{matplotlib}\xspace}

\newcommand{\molsturm}  {\texttt{molsturm}\xspace}
\newcommand{\veloxchem} {\texttt{veloxchem}\xspace}
\newcommand{\libten}    {\texttt{libtensor}\xspace}
\newcommand{\libadc}    {\texttt{libadc}\xspace}

\newcommand{\mat}{\mathbf}

\newcommand*{\mbra}[1]{\left\langle#1\middle|}
\newcommand*{\mket}[1]{\middle|#1\right\rangle}

\title{
	\adcc: A versatile toolkit for rapid development of
	algebraic-diagrammatic construction methods
}

\author{
	Michael~F.~Herbst\thanks{%
		CERMICS, \'{E}cole des Ponts ParisTech, 6 \& 8 avenue Blaise Pascal, 77455 Marne-la-Vall\'{e}e, France;
		Inria Paris, 75589 Paris Cedex 12, France;
		Sorbonne Universit\'{e}, Institut des sciences du calcul et des donn\'{e}es, ISCD, 75005 Paris, France
	},
	Maximilian~Scheurer\thanks{%
		Interdisciplinary Center for Scientific Computing, Heidelberg University, 69120 Heidelberg, Germany
	},
	Thomas~Fransson\footnotemark[2],$^{,}$\thanks{%
		Fysikum, Stockholm University, Albanova, 10691 Stockholm, Sweden
	}\\
	Dirk~R.~Rehn\footnotemark[2],
	Andreas~Dreuw\footnotemark[2]
}

\date{}
\begin{document}
\maketitle

\begin{center}
\subsubsection*{\small Article Type} Software Focus
\hfill \break
\thanks

\subsubsection*{Abstract}
\begin{flushleft}
	ADC-connect~(\adcc) is a hybrid \python/\cpp module
	for performing excited state calculations
	based on the algebraic-diagrammatic construction scheme
	for the polarisation propagator~(ADC).
	Key design goal is to restrict \adcc to this single purpose
	and facilitate connection to external packages,
	e.g., for obtaining the Hartree-Fock references,
	plotting spectra, or modelling solvents.
	Interfaces to four self-consistent field codes
	have already been implemented, namely
	\pyscf, \psifour, \molsturm, and \veloxchem.
	The computational workflow,
	including the numerical solvers, are implemented in \python,
	whereas the working equations and other expensive expressions
	are done in \cpp.
	This equips \adcc with adequate speed,
	making it a flexible toolkit
	for both rapid development of ADC-based computational spectroscopy methods
	as well as unusual computational workflows.
	This is demonstrated by three examples.
	% one providing an interactive comparison
	% between the spectra obtained for different ADC levels,
	% one showing the flexibility of \adcc
	% with respect to selecting the active
	% orbitals in the calculation of core-excited states
	% and one modelling the solvent shift of nile red.
	Presently, ADC methods up to third order
	in perturbation theory are available in \adcc,
	including the respective core-valence separation and spin-flip variants.
	Both restricted or unrestricted Hartree-Fock references can be employed.
\end{flushleft}
\end{center}

\clearpage

\renewcommand{\baselinestretch}{1.5}
\normalsize

\clearpage

%----------------------------------------------------------------------

% parts
\section*{\sffamily \Large GRAPHICAL TABLE OF CONTENTS}
\begin{center}
	\includegraphics[scale=2.5]{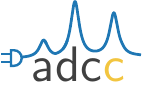}
\end{center}
\adcc: A versatile toolkit for research and teaching in computational spectroscopy
based on the algebraic-diagrammatic construction scheme for the polarisation
propagator (ADC).

% Include an attractive full color image for the online Table of Contents. It may be a figure or panel from the article, or may be specifically designed as a visual summary. You will need to upload this as a separate file during submission.

% Size: The maximum width and height are 390 pixels, and the minimum resolution is 300 dpi. Multi-panel graphs or images are strongly discouraged.

% Caption: This is a narrative sentence to convey the article's essence and wider implications to a non-specialist audience. The maximum length is 50 words, but consider using 140 characters or less to facilitate social media sharing, which can increase the discoverability of your article.

\section*{\sffamily \Large INTRODUCTION}
% Introduce your topic in around 2 paragraphs, around 750 words.

In recent years, high-level programming languages have attracted
more and more attention from computational simulation frameworks.
In the field of quantum chemistry, a multitude
of packages have emerged, which are at least partly written in languages such as
\python,~\cite{Parrish2017,Smith2018,Sun2017,molsturmDesign,Horton,PyQuante,PDynamo,Serenity,Enkovaara2011,Larsen2017}
Matlab,~\cite{Yang2009}
OCaml,~\cite{qp2}
or Julia~\cite{DFTK}.
Most prominent is \python as the scripting language of choice.
The feature sets of such packages are steadily growing and can be compared
to those of traditional quantum-chemical program packages.
In addition to features usable in practice, \python-driven packages
have also paved the way for rapid prototyping and development of new
methodologies, most notably through the \pyscf and \psifour programs \cite{Parrish2017,Smith2018,Sun2017}.
Through individual components and libraries comprising these programs,
a community-driven, open, and sustainable development can be guaranteed \cite{Parrish2017,pcmsolver}.
The aforementioned codes are freely available and encourage code contributions from external users and developers.
For improved performance, a combination of \python
with a compiled programming language, e.g., \cpp, is commonly employed.
With this approach, computationally demanding routines are handed off to \cpp,
while flexibility is maintained by exposing these routines to \python.
This exploits the strengths of each individual language:
Flexibility is maintained to the end user,
while still specialised libraries and optimised computational kernels
can be employed on the \cpp-side.
Such are commonly needed to face the unique challenges of quantum chemistry,
including high-dimensional multilinear algebra computations
in the context of post-Hartree-Fock methods.
A more detailed discussion on this hybrid \python/\cpp design
can be found in the \psifour and \psifournp publications \cite{Parrish2017,Smith2018}.
Especially the high-level reference implementations and tutorials in \psifournp
demonstrate the enormous
flexibility of a synergistic combination of \python and \cpp.

% In the context of \ADC a free-to-use / open-source / not proprietary and flexible
% framework has been lacking so far.
A family of methods that has obtained little attention in the context of hybrid
\python/\cpp program design is
the algebraic-diagrammatic construction scheme for the polarisation propagator (ADC)
\cite{Schirmer1982,Dreuw2014}. To this extent, we have developed ``ADC-connect'' (\adcc), a \python/\cpp
package for carrying out ADC calculations as well as
allowing for rapid development of ADC methods through a high-level \python interface.
Our package is a standalone toolkit that can be seamlessly connected to
any quantum-chemical host program to perform ADC methods on top of its SCF results.
Beyond the SCF, \adcc also tries to employ as much existing software as possible
for standard tasks such as tensor operations or visualisation,
allowing us to focus solely on the implementation of ADC methods.
In this manner, \adcc is agnostic of the
host program and other third-party codes of the \python ecosystem.
By keeping existing interfaces open,
we or our users do not commit to a single software stack,
but are provided full flexibility.
A large set of ADC methods is already available in \adcc and to date
\psifour~\cite{Parrish2017,Smith2018},
\pyscf~\cite{Sun2017},
\molsturm~\cite{molsturmDesign}, and
\veloxchem~\cite{Veloxchem} are fully supported
as host programs.
%The ADC and the equivalent so-called intermediate state
%representation (ISR)~\cite{Schirmer2004,Wormit2014} approaches provide
%a series of approximations for the polarization propagator~\cite{Fetter1971} based on
%diagrammatic perturbation theory and the Møller-Plesset~\cite{Moeller1934}
%Hamiltonian splitting.

\begin{figure}[h]
	\centering
	\includegraphics[width=0.97\textwidth]{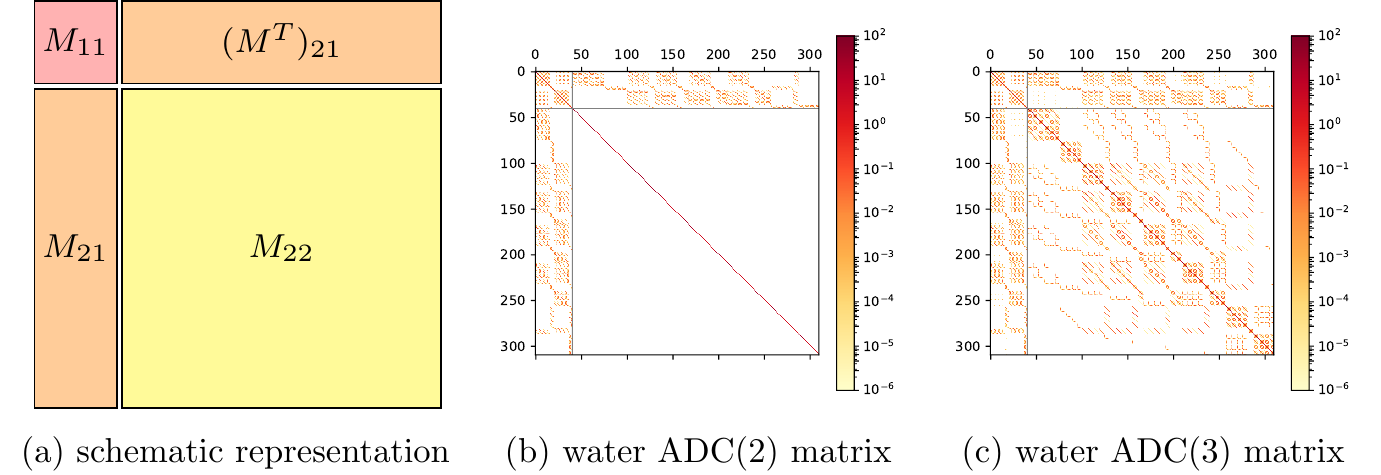}
	\caption{
		Structure of the ADC matrix.
		(a) shows a schematic representation indicating
		the singles block $M_{11}$,
		doubles block $M_{22}$ and coupling block $M_{21}$,
		(b) and (c) depict the ADC(2) and ADC(3) matrix of water
		in an STO-3G~\cite{Hehre1969} basis, respectively.
		The elements are coloured in a $\log_{10}$-scale.
	}
	\label{fig:adc_matrix}
\end{figure}
The key equation for each ADC($n$) model is the Hermitian eigenvalue problem
\begin{equation}
	\mat{M} \mat{X} = \mat{\Omega} \mat{X}, \qquad \mat{X}^\dagger \mat{X} = \mat{I},
	\label{eqn:adc_diagonalisation}
\end{equation}
where $\Omega_{nm} = \delta_{nm} \omega_n$ is the diagonal matrix of excitation energies
and $\mat{M}$ is the so-called ADC matrix. 
The matrix $\mat{M}$
exhibits a block structure, shown in Figure \ref{fig:adc_matrix}a,
where individual blocks are treated at different orders of perturbation theory.
On top of this block structure the individual blocks are sparse
(Figure \ref{fig:adc_matrix}b and c) 
which is a direct consequence of the selection rules obtained from
spin and permutational symmetry in the tensor contractions required
for computing $\mat{M}$.
Exploiting this sparsity when diagonalising
the matrix \eqref{eqn:adc_diagonalisation}
is a key step to make ADC tractable for relevant systems.
In this regard, \adcc follows the conventional approach~\cite{Dreuw2014,Wormit2014}
to use contraction-based~\cite{molsturmDesign}, iterative
eigensolvers, such as the Jacobi-Davidson~\cite{Davidson1975}.
All tensor operations in the required ADC matrix-vector products
are performed on block-sparse tensors.
In this setting, the computational scaling of ADC(2) is given as $O(N^5)$
where $N$ is the number of orbitals,
whereas ADC(2)-x and ADC(3) scale as $O(N^6)$~\cite{Dreuw2014}.
%Solving the secular equation gives access to excitation energies and transition moments.
%The eigenvectors $\mat{X}$ can be used to calculate excited state and state-to-state
%transition properties by virtue of the ISR approach.
%In addition, this approach offers a convenient route to the calculation of linear and
%non-linear molecular response
%properties~\cite{Trofimov2006,Knippenberg2012,Fransson2017,Rehn2017a}.
%Further modifications of the ADC scheme are the so-called core-valence separation~(CVS)%
%~\cite{Cederbaum1980,Trofimov2000,Wenzel2014b,Wenzel2014a,Wenzel2015}
%for the description of core-excited states,
%the spin-flip~\cite{Lefrancois2015,Lefrancois2016,Lefrancois2017}
%ansatz for few-reference problems, multi-reference ADC~\cite{mradc2018}
%and four-component ADC~\cite{reladc2014}.
%Despite the growing popularity and numerous applications%
%~\cite{adcdynxas2016, reladc2018, adcwaterrixs2018}, none of the aforementioned
%packages provide a comprehensive implementation of ADC methods nor
%the possibility of rapid prototyping and
%\python-driven development for ADC methods.
%
These computationally demanding procedures are implemented in \cpp
through the parallelised \libten library \cite{Libtensor}
and made available to the \python layer.
In this manner, \adcc achieves comparable performance as a
\cpp-only implementation~\cite{Wormit2014}~(cf.~Table SI-1).
Still, all data is readily accessible from \python,
e.g., to perform post-processing in \numpy \cite{Walt2011}
or \matplotlib \cite{Matplotlib}.
Consequently, our design enables a very flexible computational workflow
and interactive usage in Jupyter notebooks or IPython shells \cite{IPython,Jupyter}.
This will be demonstrated by several examples.
Additionally, exposing linear algebra operations of
\cpp tensor objects to the \python layer
made it possible to implement numerical procedures,
e.g., iterative solver schemes completely in \python.
As such, \adcc provides all the necessary building blocks for development of complex
workflows or novel approaches of ADC that can be arbitrarily assembled
and extended on the \python layer.
Together with the comprehensive documentation
available online (\url{https://adc-connect.org}), the
barrier between ``users'' and ``developers'' of \adcc is minimal.
\adcc is freely available on \url{https://github.com/adc-connect/adcc}.
All figures and tables of the paper can be reproduced using the scripts
and details of the supporting information (SI-4).

% This not only reduces the feedback loop between sending a command
% and receiving a result, but also allows to inspect results directly at the spot.
% This is great for teaching, for development, and also for investigating
% a computational result.

% This means that the barrier to include practically motivated useful
% workflow and features into the package is substantially lowered.

% A remarkable advantage of scriptable, high-level interfaces for computations
% is that they lower the traditional barrier between a ``user'' of a package
% and a ``developer'', since a more complicated computational workflow
% can be considered a ``program'' on its own. 

% This minimizes the feedback loop during method development since no code compilation is necessary.

% 
% \subsection{Why \python?}
% \emph{Scriptable, high-level approach to computational spectroscopy.}
% 
% Without a doubt nowaday computational spectroscopy procedures
% are complex. Even within the ADC context, there are many questions
% to be kept in mind, which have an influence on the workflow.
% One example being solvent models or embedding theories,
% another one being the generation of a guess excitation
% using a cheaper level of theory (see solver
% cascade example \ref{sec:ex:comparison}).
% \MS{the previous paragraph can be dropped (duplicate): flexible, complex workflow already explained

The remainder of the paper is structured as follows:
The next Section discusses design and structure of \adcc
with emphasis on the computational workflow and the
integration with external packages.
Examples how \adcc can be used in practice are provided thereafter,
followed by a short review of the currently
supported feature set of \adcc.

\section*{\sffamily \Large DESIGN AND KEY COMPONENTS}
\label{sec:design}

\section*{\sffamily \Large Design goals}
\label{sec:goals}

With \adcc, we aimed for a flexible library for rapid
ADC method development that seamlessly integrates into the existing \python ecosystem.
Consequently, secondary goals for the design philosophy of the code arise
which are briefly outlined in the following.

\begin{enumerate}
	\item \emph{Build on established software.}
		Instead of designing a complete quantum-chemical program,
		\adcc focuses only on ADC methods and their numerical procedures.
		For most other aspects, \adcc relies on established third-party packages,
		which allows to reuse many years of development effort and bug fixing.
		This makes \adcc a light-weight package
		to be easily integrated in more complex
		computational workflows.
	\item \emph{Open interfaces for reproducible and sustainable science.}
		The SCF interface of \adcc is deliberately kept simple
		for easiest connectability to an existing SCF codes.
		This has two important consequences:
		(1) It makes \adcc sustainable and maintainable,
		because in case an SCF code became unavailable
		in the future, a replacement could be easily integrated.
		(2) Results can be verified and reproduced
		across different SCF implementations,
		reducing the chance of building on top of
		wrongful assumptions and introducing accidental dependencies.
	\item \emph{Good compromise between performance and code complexity.}
		Given the asymptotic scaling up to $O(N^6)$ of ADC methods,
		performance aspects cannot be ignored
		in a practically useful implementation.
		The total time required to achieve
		a scientific outcome in computational sciences, however,
		clearly goes beyond just the runtime of simulations.
		Much rather, the time needed to install,
		setup, and familiarise oneself with a framework also matters.
		In case additional features are to be added to the code,
		the time effort to understand the code base,
		to implement, debug, and test also plays a pivotal role.
		Unfortunately, achieving peak performance typically
		has a negative impact on readability and usability,
		such that a balance needs to be found.
		In \adcc we have thus chosen a design,
		where the workflow and the numerical schemes
		are completely controlled from high-level \python code
		with only selected
		computationally demanding parts implemented in \cpp.
		This allows \adcc users both to quickly experiment with workflows
		or numerical routines and to treat problems of practical relevance
		in a single toolkit.
	\item \emph{Low barriers between users and developers.}
		With our toolkit we not only want to allow for
		use of provided, predefined functionality,
		which already exists inside \adcc.
		By providing detailed documentation and readable code,
		we want to encourage users to
		become active developers, e.g., by extending our workflows,
		beyond anything we as the \adcc developers could ever imagine.
		It is our hope that such an open platform will
		% will encourage contributions from our users back into the main code,
		lead to a community-driven improvement of \adcc in particular
		and computational spectroscopy in general.
\end{enumerate}

\section*{\sffamily \Large Demand-driven computation in \adcc}
\label{sec:demand_driven}

\begin{figure}[h]
	\centering
	\includegraphics[width=0.4\textwidth]{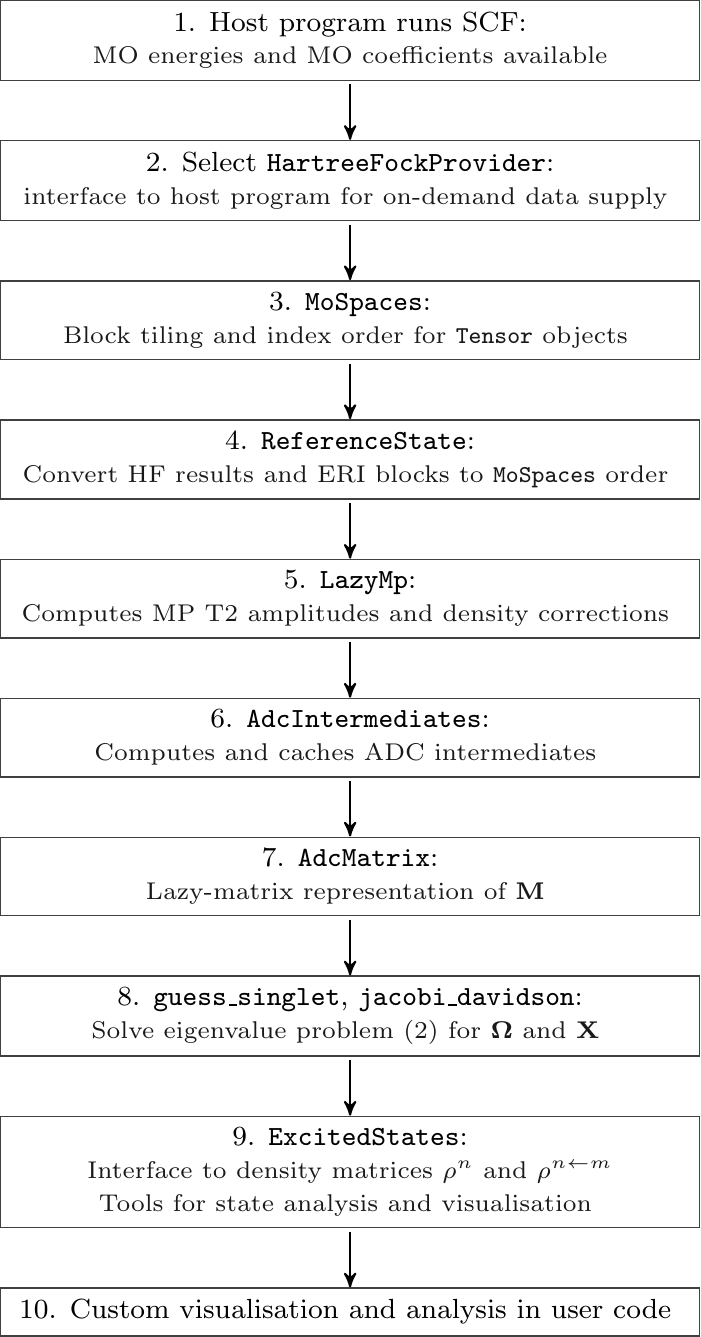}
	\caption{
		Schematic of an ADC calculation using \adcc.
		Steps 2 to 9 take place inside \adcc,
		whereas steps 1 and 10 are executed in the host program or user code.
		Functions or classes from the \adcc library are
		marked with \texttt{teletype} font.
	}
	\label{fig:adcc_flow}
\end{figure}

The general flow and main computational
tasks of an ADC calculation with \adcc is shown in
Figure \ref{fig:adcc_flow} with
classes or functions from \adcc given in \texttt{teletype}.
The term ``host program'' refers to the (third-party)
program environment from which the ADC calculation has been started.
In practice, this is the code that yields the Hartree-Fock SCF ground state
and provides access to an integral library for obtaining
the antisymmetrised electron-repulsion integral~(ERI) tensor
or operator integrals (e.g., electric dipole operator integrals).
Steps 2--4 are setting up the necessary scaffold to import host-program-specific
data into \adcc.
Afterwards, the main bookkeeping classes of \adcc are constructed in steps 5--7,
i.e., quantities of the Møller-Plesset~(MP) ground state,
recurring parts of the ADC working equations (intermediates)
\cite{wormit2009diss}, and a \textit{lazy matrix} \cite{molsturmDesign,Herbst2018Phd}
representation of the ADC matrix.
Of note, this \texttt{AdcMatrix} class collects \emph{all} MP results,
intermediates, and expressions for the working equations of a particular ADC method.
The ADC eigenvalue problem is then solved in step 8, and the results
are wrapped by an \texttt{ExcitedStates} object,
see the next Section for details.

With the workflow schematic of \adcc in place,
we will now explain how \adcc efficiently implements this complex
computational procedure.
Here, the main challenge is that individual computational steps can explicitly
or implicitly become mutually dependent. This not only applies to ADC, but
also to related methodologies with similar computational stages.
The ADC intermediates for example
(step 6 in Figure \ref{fig:adcc_flow}) require the MP(2) T2 amplitudes.
Therefore, step 6 obviously depends on step 5.
In turn, a naive sequential implementation requires step 5 to know that
the T2 amplitudes will be needed in step 6
in order to decide whether these quantities are to be computed or not.

One way to resolve this circular dependency is to compute every
quantity at every step,
regardless of whether it is used later or not. 
This approach, however, is rather inefficient.
Another option is to inspect the computational parameters
at the beginning of the \adcc run and schedule computations for each step.
While this ``schedule'' approach works reasonably well for simple workflows,
it inevitably leads to a combinatorial explosion in the required
bookkeeping logic if the number of parameters increases.
Already at the current stage, \adcc has about 40 different code execution paths
through the workflow in Figure \ref{fig:adcc_flow}
with varying amounts of work required at each step.
% This number is likely going to increase in the future
% as more methods are implemented in \adcc.

For this reason we have taken a different approach:
Upon initialisation of data structures, such as
\texttt{ReferenceState}, \texttt{LazyMp}, or \texttt{AdcIntermediates},
these objects are empty. For example, constructing \texttt{LazyMp}
does not lead to the computation of the T2 amplitudes at this very instance.
Only once the T2 amplitudes are needed for the first time,
e.g., by computation of the
ADC(2) intermediates, they are computed and cached in memory.
Thus, the first demand for a specific quantity drives its computation.
For this reason we have termed this strategy ``demand-driven''.
This idea is heavily inspired from a concept called \textit{lazy evaluation}
in programming language theory~\cite{Hughes1990},
where any expression in the source code is only ever evaluated
once its outcome is needed.
% Note that the caching logic for demand-driven computation requires additional code
% beyond the expressions for computing, e.g., the T2 amplitudes.
% In contrast to the schedule approach,
% it is, however, easy to keep such code close to the
% algorithmic expression it guards.
In \adcc, the caching check is typically an \texttt{if}-statement
directly enwrapping the computation.
This has two important advantages.
Firstly, it makes the code responsible for computing one quantity
self-contained,
% and decoupled from the remainder of \adcc,
which lowers the code complexity.
Secondly, it prevents schedule logic and computational algorithm to get out of sync.
% If, for example, the computation of the ADC(2) intermediates were changed
% and as a result the T2 amplitudes were no longer needed for the ADC(2) intermediates,
% in a demand-driven computation the code execution path would
% automatically adapt.
% In case of the schedule strategy, however,
% the schedule logic would need to be manually adapted.
Thus, a major advantage of the demand-driven computation is that
one simply cannot forget to request the computation of a particular
quantity in a previous step or forget to remove this request.
This decouples code entities and makes \adcc automatically 
choose the path of least computational load.

%\todoil{In experiments setting \texttt{export MALLOC\_MMAP\_THRESHOLD\_=1024}
%shows an easier-to-interpret memory profile, but the calculation is slower.}
\begin{figure}[ht!]
	\centering
	\includegraphics[width=\textwidth]{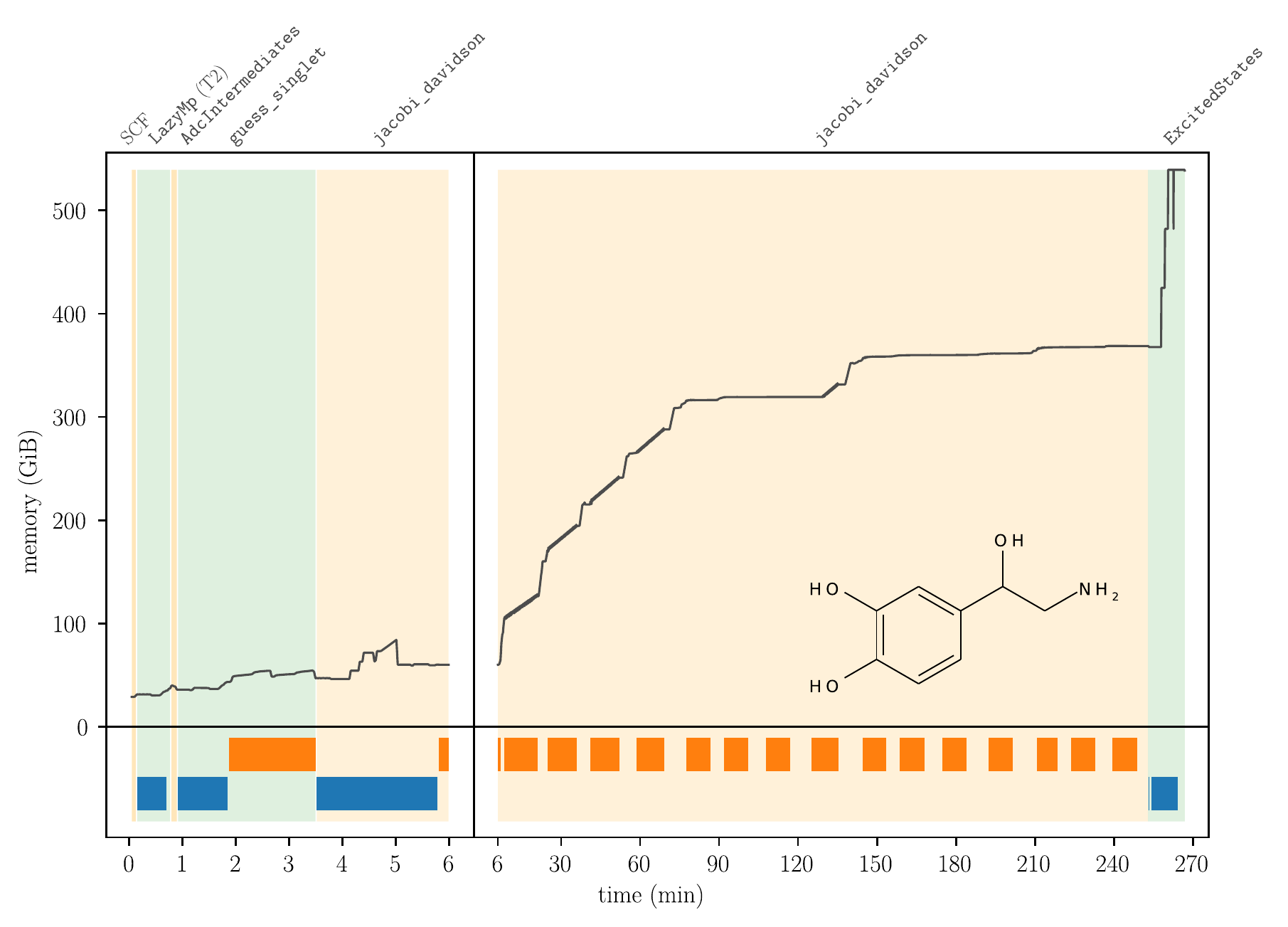}
	\caption{
		Memory and time profile of an ADC(2) calculation
		for the lowest 10 singlet excitations
		of noradrenaline in a 6-311++G** basis~\cite{6311Gstarstar}.
		The green and yellow background indicate
		the time spent in functions or classes of \adcc,
		which are labelled in \texttt{teletype} font.
		The orange and blue bars at the bottom
		indicate the activity of the \texttt{AdcMatrix} class
		and the time spent importing the electron-repulsion
		integrals, respectively.
		%The blue ERI import bar shows from right to left
		%the import of the ERI blocks (oovv, ovov, ovvv, oooo, vvvv),
		%where the oooo import is barely visible.
		Details about the computational hardware
		can be found in the supporting information,
		Section SI-3.
	}
	\label{fig:memtrace}
\end{figure}
To illustrate the demand-driven workflow
in practice, Figure \ref{fig:memtrace} shows
the memory and time profile for a
computation of ten singlet excited states
of noradrenaline at the ADC(2)/6-31++G** level~\cite{6311Gstarstar}.
The time spent in selected classes and functions of \adcc
is indicated by the alternating yellow-green background,
and the memory size of the executing \python process as detected
by the operating system is plotted in grey.
Of note, the memory profile not only
depends on the behaviour of \adcc,
but also on the memory management
of the \python interpreter and the C library (in this case \texttt{glibc}).
The time axis is split into two segments
with the first resolving the first six minutes of the calculation to higher detail.
The lower part of Figure \ref{fig:memtrace} shows two bars of orange and blue blocks,
which highlight, respectively,
the time spent inside \texttt{AdcMatrix}
for computing the ADC working equations
and the time needed for importing the ERI tensor.

One notices the ERI tensor not to be imported at once,
but at four distinct times in the profile.
Each time interval corresponds to computing one block of the ERI tensor
in the host program via an AO-to-MO transformation
followed by importing it into the \texttt{ReferenceState}.
Through the demand-driven design of \adcc,
one can easily follow the order of the block import, summarised in Table \ref{tab:eri_import}.
Note that the objects requiring import of a new ERI block often also
depend on previously imported blocks that are already cached.

% \MS{original text}
% Denoting by $o$ the set of all occupied orbitals
% and by $v$ the set of all virtual ones,
% the first import is the $\mbra{oo}\mket{vv}$ block.
% This block is needed to compute the T2 amplitudes in \texttt{LazyMp}
% which are required in turn for the ADC(2) intermediates
% in \texttt{AdcIntermediates}.
% The second import around \unit[1]{min} is the $\mbra{ov}\mket{ov}$ block,
% required for the ADC(2) diagonal, which in turn
% is needed to construct the guesses in \texttt{guess\_singlet}.
% Next, around \unit[3.5]{min}, the $\mbra{ov}\mket{vv}$
% as well as the $\mbra{oo}\mket{ov}$
% blocks are imported, which are needed in the ADC working equations
% on top of the previously imported blocks and the ADC intermediates.
% Finally the $\mbra{oo}\mket{oo}$ and $\mbra{vv}\mket{vv}$
% blocks are imported around \unit[250]{min} for the computation
% of the transition densities $\rho^{n \leftarrow 0}$
% and the property calculations in the \texttt{ExcitedStates} class.
% Note that the two blocks are not imported at the same time,
% but the import of the $\mbra{oo}\mket{oo}$ takes only very little time
% (around 20 seconds) and is thus barely visible in figure \ref{fig:memtrace}.

\begin{table}[h]
	\centering
	\caption{
		Breakdown of the demand-driven import of ERI blocks
		for the noradrenaline calculation of Figure \ref{fig:memtrace}.
		The first demand of a block is indicated by the operations
		of the rightmost column.
		$o$ refers to occupied orbitals and $v$ to virtual orbitals.
		Multiple blocks in one row are imported sequentially.
	}
	\label{tab:eri_import}
		\begin{tabular}{cl@{\qquad$\leftarrow$ }c@{ $\leftarrow$ }c}
			\toprule
			time (min) & ERI block & \multicolumn{2}{c}{first demand} \\
			\midrule
	0	& $\mbra{oo}\mket{vv}$ & \texttt{LazyMp} T2 & \texttt{AdcIntermediates} \\
	1	& $\mbra{ov}\mket{ov}$ & \texttt{AdcMatrix.diagonal} & \texttt{guess\_singlet} \\
	3.5	& $\mbra{ov}\mket{vv}$, $\mbra{oo}\mket{ov}$ & \texttt{AdcMatrix.matvec}
		& \texttt{jacobi\_davidson} \\
	250	& $\mbra{oo}\mket{oo}$, $\mbra{vv}\mket{vv}$ & transition densities
		& \texttt{ExcitedStates}\\
		\bottomrule
		\end{tabular}
\end{table}

The complete chain of imports and computations in Table \ref{tab:eri_import}
has actually been triggered by requesting only three things directly from \adcc,
namely (1) the ADC(2) guesses via \texttt{guess\_singlet},
(2) the converged ADC(2) excitation vectors from \texttt{jacobi\_davidson}
and (3) the computation of the oscillator strengths for these states
via \texttt{ExcitedStates}.
The other computations including the import of the ERI tensor blocks
were implicitly driven by this initial demand.
If this demand is modified,
e.g., by performing a computation employing the CVS, frozen-core~(FC)
or frozen-virtual~(FV) approximations,
the ERI blocks not required by the respective approximations,
will never be computed in the host program during the ADC calculation.
Still, dropped blocks can be requested
via the \texttt{ReferenceState} in user code,
meaning that any additional demand on top of the ADC calculation
will be satisfied as well.
This is of great advantage during debugging and in order to
extend features of \adcc in user code.

Typically the import of an ERI tensor block
leads to an increased memory usage, since tensor data is generated
with the AO-to-MO transformation in the host program.
As discussed, in \adcc the ERI import is automatically delayed
for as long as possible.
This implies that the allocation of the ERI tensor memory is delayed as well.
As a result, the peak memory usage for the noradrenaline calculation
in Figure \ref{fig:memtrace} is only obtained at the very end of the
calculation, namely during the property calculation,
once the $\mbra{vv}\mket{vv}$ block has been imported.
This memory profile implies that \adcc will run out of memory as late as possible.
In other words it will still finish intermediate work,
from which the calculation could be restarted on a node with more RAM.
Note that at the moment \adcc does not yet implement
any form of checkpointing, however.

Regarding the memory requirements of \adcc it should be noted
that a number of expert interfaces to influence the default memory footprint exist.
Using a custom \texttt{CachingPolicy}, for example, one may
disable the storage of a particular quantity,
such as the ADC intermediates,
at the expense of recomputing it on the next demand.
Additionally explicitly purging intermediates or selected ERI blocks from memory
is possible from our \python interface.

\section*{\sffamily \Large Structure and mix of programming languages in \adcc}
\label{sec:design_structure}
\begin{figure}[h]
	\centering
	\includegraphics[width=\textwidth]{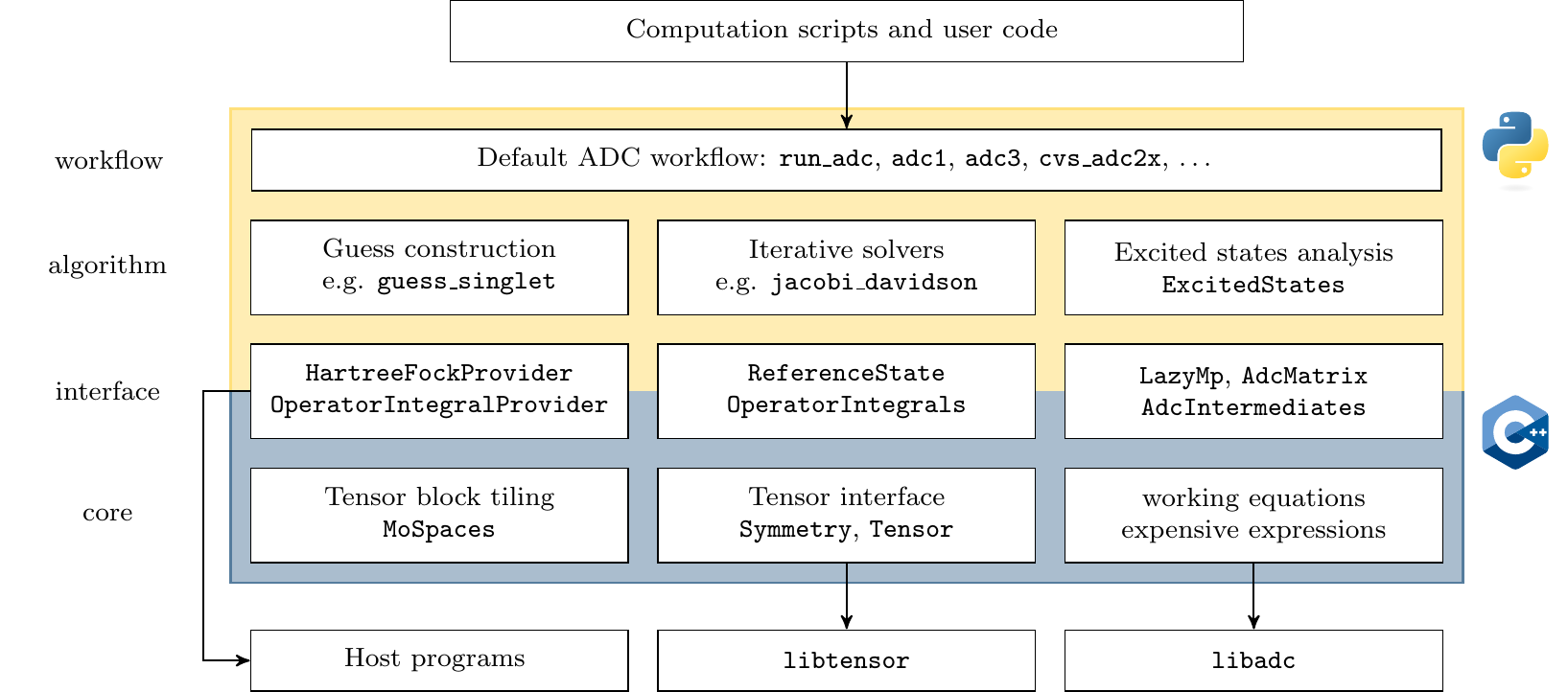}
	\caption{
		Code structure of \adcc and interfaces
		to user scripts, external host programs
		or the \cpp libraries \libten~\cite{Libtensor}
		and \libadc~\cite{Wormit2014}.
		Shown inside the yellow-blue box are the four layers of \adcc
		with key classes and functions marked in \texttt{Teletype} font.
		The background of the box indicates the predominant
		programming language employed in the respective layer,
		yellow for \python and blue for \cpp.
	}
	\label{fig:structure}
\end{figure}
With the \adcc workflow and the concept of demand-driven
computations in place, we will now explain the actual structure and
building blocks of \adcc in detail, together with the hybrid \python-\cpp
programming approach.

Figure \ref{fig:structure} shows the code structure of \adcc
along with interfaces to external codes or user scripts
as indicated by arrows leaving or entering the \adcc box.
Inside the box, the colours indicate the predominant
programming language used to provide the respective features.
Going from top to bottom,
the first layer is the \textit{workflow layer}. It contains
a default ADC workflow for setting up the \texttt{AdcMatrix}
% corresponding to the requested ADC method (redundant)
and subsequently calling the
functions of the \textit{algorithm layer}
to solve the ADC eigenvalue problem \eqref{eqn:adc_diagonalisation}.
Main ingredients of this second layer
are the guess functions, the \python-based implementations
of the iterative solver algorithms
and the \texttt{ExcitedStates} class to analyse the obtained results.
Next, the \textit{interface layer} contains the datastructures
responsible for the ADC working equations and their requirements,
including the \texttt{HartreeFockProvider}
and \texttt{OperatorIntegralProvider} interface classes for each host program.
As discussed, most computation in \adcc happens in this layer,
demanding the computation or import of prerequisites as needed.
For performing these tasks,
the interface layer accesses helper functionality from the
bottommost part of \adcc, the \textit{core layer}.
This layer contains the \texttt{MoSpaces} class
for defining the tensor tilings as well as
the \texttt{Tensor} and \texttt{Symmetry} classes,
which expose slightly amended primitive tensor operations
from the \libten~\cite{Libtensor} block-sparse tensor library to \adcc.
It also contains the interface code to the \libadc library,
which implements the ADC working equations and other expensive expressions
in \libten syntax.
Both these libraries are written in \cpp,
such that implementing the remainder of the core layer in \cpp
turned out to be most practical.
The \libadc library has been previously distributed
as part of the \adcman~\cite{Wormit2014} module of the Q-Chem
quantum chemistry package \cite{Shao2015}.
As part of the interface to \libadc, the core layer further makes use of the
hierarchical storage datastructures of the ctx library~\cite{ctx}
for organising the data required for evaluating the ADC equations.

%
% Mixed languages and balance between performance
%    - Compromise between performance and code complexity
%       python and C++ with numerical algorithms coded in python!
%

% We already noted
% that the most time-consuming part of an ADC calculation is the evaluation
% of the ADC working equations,
% compare the blue bars at the bottom of figure \ref{fig:memtrace}.
% For this reason performance is most critical in all parts of code
% related to these hot spots of the computation.

% Most of the time-consuming work of \adcc is directed to the \libadc library
% and the \libten library,
% which provides automatic node-level parallelism
% for the required tensor contractions. 
% I (MS) put the \cpp stuff above, where the libraries are mentioned first (on-demand ;) )

On the contrary, the interface layer,
as indicated by the colour-coding in Figure \ref{fig:structure},
is partly written in \python and partly in \cpp.
The divide between these languages is not at all clear-cut:
Much rather some functions and datastructures are implemented
\emph{using both languages}.
Using Pybind11~\cite{pybind11} the respective \cpp functionality is exposed to \python,
where it can be accessed and extended without duplicating code.
In this way, the interface to performance-critical portions
of \adcc in the core layer is implemented
in \cpp and the remaining aspects may fully exploit
the language features \python has to offer.
One example where this approach turned out to be very
useful are the \texttt{HartreeFockProvider}
interfaces, which on the one hand need to interact with
a third-party host program,
whose interfaces we as \adcc developers cannot control,
on the other hand it still needs to provide the generated data
to the \libten library.
For the former aspect, a dynamic language like \python is extremely handy,
while for the latter, low-level memory access is required
and thus a language like \cpp is much more suitable.

The algorithm layer itself is also an example for \python code,
which extends the \cpp core.
By proper Pybind11-wrapping of key classes,
such as the \texttt{MoSpaces}, \texttt{Tensor}, and \texttt{Symmetry}
classes, the algorithm layer
can directly configure and access
\emph{all} the functionality of the core layer albeit being written in \python.
This includes raw tensor operations such as addition, multiplication
or tensor contraction.
Provided that the tensor tiling in \texttt{MoSpaces} has been setup appropriately,
these operations automatically take symmetry
into account and are parallelised.
In this way all relevant numerical schemes of ADC,
such as the iterative diagonalisation inside the \texttt{jacobi\_davidson},
can be implemented purely in \python.
Certainly, a multitude of calls between
\python and \cpp are required which are associated with a runtime overhead.
Since the time spent for computing tensor operations in \libadc, \libten,
and the core layer dominates the overall runtime, such that
call time overhead as well as the general performance penalty
of \python are completely negligible.
Compared to a \cpp-only implementation of the algorithm,
however, flexibility is gained through \python.
This enables more involved
or highly problem-specific numerical schemes,
which can be developed, tested, and implemented with much reduced effort.

Keeping the implementation of iterative solvers and guesses aside,
the algorithm layer also contains the \texttt{ExcitedStates} class.
This class is returned to the user once the ADC calculation in \adcc has finished
and holds key results of the calculation
such as the excitation vectors or the excitation energies.
From these, state densities $\rho^n$,
transition densities $\rho^{n\leftarrow m}$,
and other properties can be computed on demand.
The returned quantities are in fact \texttt{Tensor} objects,
allowing the user to post-process them directly using
the tensor operations of \libten.
Alternatively, these objects may be converted to and from dense \numpy arrays
to allow full integration with the usual
\numpy/\scipy ecosystem~\cite{Walt2011,scipyWeb}.
More details about customisable post-processing can be found in
the example Section and on the \adcc website.
% This can be used to implement post-processing steps,
% such as the computation of natural transition orbitals
% by decomposing the transition density matrix with
% \texttt{scipy.linalg.svd}
% in only a few lines of \python code on top of \adcc.
% Naturally, such density matrices and excitation vectors
% may also be passed to third-party molecular simulation software
% to implement excited states methods derived off plain ADC.
% One example is the inclusion of solvent models%
% ~(see section \ref{sec:ex:nile_red}).
Some conventional analysis and visualisation techniques,
such as the plotting of an empirically broadened excitation spectra
are directly available by calling the \texttt{plot\_spectrum} function of the \texttt{ExcitedStates}.
These functions integrate well with existing \python infrastructure,
in this case \matplotlib~\cite{Matplotlib}.
% \MS{users should check this out on the website}
% For example, for the color of the plotted spectra,
% the axis labelling or the canvas size \adcc automatically chooses a default
% preset giving reasonably-looking results.
% Still the user is able to use matplotlib functionality
% to extend or modify these presets.
An example for the spectra plotting of \texttt{ExcitedStates}
can be found in the examples.
Such tools to quickly visualise results are key
for reducing the feedback loop when working with \adcc from,
e.g., a Jupyter notebook~\cite{Jupyter} or for educational purposes.

A user who is new to \adcc will start to interact with the library
mostly via the workflow layer,
calling the predefined ADC procedure
it exposes via the \texttt{run\_adc} \python function.
More method-specific functions with slightly varying presets are available
for individual ADC methods,
for example \texttt{adc1}, \texttt{cvs\_adc3}, \ldots.
These functions provide only limited capability for customisation,
passing parameters such as the subspace size or
the structure of the core space (for core-valence separation)
to the rest of the library.
For more advanced use cases, like the ones mentioned above,
they may then leverage the \texttt{ExcitedStates} class
or construct deviating ADC workflows
building on the \python primitives of the algorithm and interface layer.
Simulation procedures resulting from this process
\emph{are} already developments (in \python code),
which could potentially be integrated back into the library in the future.
In this way the original user has smoothly become a developer of \adcc.
Even for yet deeper modifications one may stay in high-level \python code,
being able to tinker with advanced aspects such
as the setup of the ADC guesses or the numerical procedure
to diagonalise the ADC matrix.
% \MS{Probably drop the next paragraph? We do not allow for this at the moment anyways...}
% Only for going beyond the interface layer,
% which is in fact only necessary for implementing fundamentally new ADC methods,
% the \cpp core layer needs to be modified.
% In this case the similarity in structure between the
% \cpp and \python part of \adcc should facilitate the transition
% to this lower layer as well.
Overall, it is our hope that the sketched structure of \adcc
allows to motivate users to become developers as well.
% , first at the \python
% level and over time, building on their familiarity with the code base,
% to be able to even perform deep modications of the code at the core level.

Related to the aspect of obtaining a sustainable base of users and developers
is the question of a sustainable software stack.
For this it is important to (1) build on top of software, which is already
established and thus unlikely to disappear
and (2) to stay sufficiently flexible
to be able to swap components if this may still become necessary.
We achieved this by building on top of two actively developed \cpp libraries,
\libten and \libadc \cite{Libtensor,Wormit2014}, and
% In \adcc the result of trying to keep to these two principles is
% that the core layer
% almost entirely consists of code interfacing to \libten and \libadc,
% two libraries, which have been developed actively
% for many years~\cite{Libtensor,Wormit2014} as part of Q-Chem~\cite{Shao2015}.
abstracting from these libraries explicitly through
the \texttt{Tensor} and \texttt{Symmetry} classes.
In this way, clearly defined entry points from \adcc
to these third-party codes are defined, such that other tensor libraries
could be supported in the future as well.
Similarly, with respect to the host programs
for supplying integral and SCF data,
\adcc supports multiple SCF codes out of the box.
% \MS{already in introduction and design goals}
% meaning that the user is able to transparently start an ADC calculation
% from a third-party SCF result without needing to explicitly worry about
% converting or passing data between the host program and \adcc.
% Right now four codes are supported to this extend, namely
% \pyscf~\cite{Sun2017}, \psifour~\cite{Parrish2017,Smith2018},
% \molsturm~\cite{molsturmDesign} and
% \texttt{veloxchem}~\todo{Reference}.
As will be explained in more details in the next Section, %\ref{sec:import},
the SCF interface expected by the \texttt{HartreeFockProvider}
has been designed to be easily fulfilled,
such that support for further host programs can be added with ease.

\section*{\sffamily \Large HF interface and ERI data import}
\label{sec:import}

To start an ADC calculation, \adcc requires two kinds of data from the host program.
Firstly, the obtained SCF results,
such as the molecular orbital energies, coefficients and occupation vector
as well as the Fock matrix and the (antisymmetrised) ERI tensor.
Secondly, metadata such as the SCF convergence threshold
and whether or not a restricted SCF procedure was employed.
Only optionally,
if the computation of properties such as dipole moments
and oscillator strengths is desired,
\adcc requires further the total molecular charge,
the nuclear dipole moment
and the electric dipole integrals in the atomic orbital basis.
We will only review key design aspects about the interface between \adcc and
host programs in this section, the complete documentation
can be found in the supporting information SI-1 and SI-2
or online under \url{https://adc-connect.org/q/hostprograms}.

At the moment data can be supplied to \adcc in three ways.
The most straightforward implementation only requires one to prepare
a \python dictionary, which supplies the above quantities as
either primitive \python data types or as \numpy arrays.
The antisymmetrisation of the ERI tensor may either be performed in \adcc
or the antisymmetrised tensor may be supplied directly.
This dictionary is then passed to \texttt{run\_adc} or any
other method from the workflow layer to start an ADC calculation.
For this type of interface the key focus was simplicity rather than efficiency.
Therefore, \adcc makes no attempt to exploit any kind of symmetries for
the Fock matrix or the ERI tensor.
Even for restricted SCF results, all four spin blocks of the Fock matrix
need to be passed,
including the $\alpha\beta$ and $\beta\alpha$ blocks, which are always zero.
Naturally this leads to a huge memory overhead and thus
allows this interface only to be used for small test calculations.
The second option is a variant of the dictionary-based interface,
where the data is supplied from an HDF5 file~\cite{HDF5Manual}.
This version integrates more closely into the demand-driven workflow,
such that \adcc will only read those parts of the tensors from disk, which are needed.

The best performance for the data import, however,
is achieved through an implementation of a \texttt{HartreeFockProvider} and a
corresponding \texttt{OperatatorIntegralProvider}
specific to the host program.
As the names suggest, the former is responsible for all
HF-related quantities and the latter for
integrals such as the electric dipole operator integrals.
Both classes allow to integrate host-program-specific routines,
e.g., for the AO-to-MO transformation, into the demand-driven workflow of \adcc.
This requires the definition of about 20 functions,
of which most are trivial and only require the user to return plain data.

The integration with the import of the ERI tensor, however,
is more involved and the design will be briefly discussed.
Once the first demand towards an ERI tensor block
triggers the import process inside the \texttt{ReferenceState} class,
the tensor tiling and tensor symmetry is already known
and can be used to deduce the minimally required subset of tensor elements
to fully represent the complete block in question.
A block-sparse tensor library such as \libten
only holds these elements in memory and
consequently only these need to be filled with data by the host program.
In the import code one thus only loops over
symmetry-reduced chunks of the ERI tensor
and requests the respective tensor data from the \texttt{HartreeFockProvider}.
This is done by passing it an index range, some details about the memory alignment
and a data pointer with the memory location to
place the generated ERI elements into.
If this memory pointer can be directly employed as output memory
inside the AO-to-MO routine of the host program,
the ERI tensor import operates without a single copy.
The import is implemented inside the \cpp part of \texttt{ReferenceState}
as well as the core layer of \adcc.
On this low level a link between \adcc and a host program
can be achieved directly by inheriting from a \cpp abstract base class.
One may also implement a \texttt{HartreeFockProvider} in \python,
since we employ Pybind11~\cite{pybind11}
to allow \python classes to overwrite the \cpp base class.
At this level \adcc also uses \numpy arrays to conveniently
hide the details of the memory pointer without an additional copy.
An implementation of a \texttt{HartreeFockProvider}
from \python thus has the full flexibility of \numpy and \python
to interact between host program and \adcc, but at no additional expense.

Since all three interface approaches are \python-based,
they can be mixed. This is useful for adding initial
support of a new host program,
since one can start with a rapid prototype
using the dictionary-based approach.
Based on this the functions of the \texttt{HartreeFockProvider}
can be implemented step by step,
verifying correctness along the way.

\section*{\sffamily \Large EXAMPLES}
\label{sec:examples}
%%%%%%%%%%%%%%%%%%%%%
%--   Section A   --%
%%%%%%%%%%%%%%%%%%%%%
\section*{\sffamily \Large Comparison of ADC methods}
\label{sec:ex:comparison}

\begin{figure}[h]
	\centering
	\includegraphics[width=0.98\textwidth]{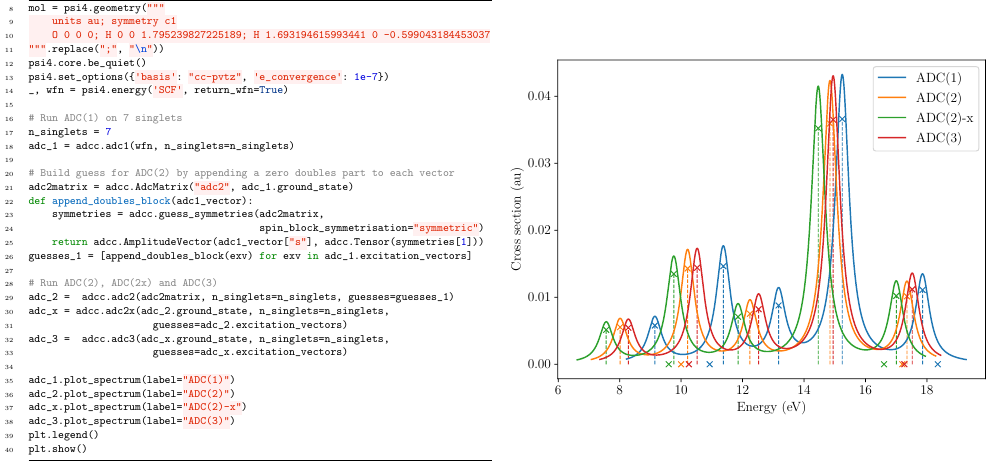}
	\caption{
		\python script computing the seven lowest-energy
		singlet excited states of water
		in a \mbox{cc-pVTZ} basis~\cite{Dunning1989}
		at ADC(1), ADC(2), ADC(2)-x and ADC(3) level
		and resulting excitation spectrum.
		The procedure uses
		the respective lower level of theory
		as a guess for the next computation.
		Spectra are broadened empirically
		with a Lorentzian
		with width parameter $\gamma = 0.01$ atomic units
		and shown in the same colour as the computed
		excitation energies and cross sections,
		which are marked by a cross.
	}
	\label{fig:SpectrumSolverCascade}
\end{figure}

A frequent task in benchmarking
is to compare the result of multiple levels of theory on the same system.
Using \adcc, different ADC methods can be used upon the same system
with concise \python code, shown in Figure \ref{fig:SpectrumSolverCascade}
alongside the resulting spectra.
In lines 8 to 14,
the script prepares a restricted Hartree-Fock reference
of water in \psifour using a \mbox{cc-pVTZ} basis~\cite{Dunning1989}.
On top of this the \texttt{adc1}, \texttt{adc2}, \texttt{adc2x}
and \texttt{adc3} functions of \adcc
perform the respective ADC method on top.
In each case, the \texttt{excitation\_vectors} of the \texttt{ExcitedStates} object
of the lower level of theory are used as guesses,
which for the case of employing ADC(1) results in ADC(2)
requires to append a zero doubles part of the appropriate singlet symmetry
in lines 22 to 26.
For starting the ADC calculation
only the first invocation in line 18 makes reference
to the \texttt{wfn} object containing the HF reference
in the form of an interface for obtaining molecular orbital coefficients,
the ERI tensor, operator integrals and so on.
All other calculations start directly from a \texttt{LazyMp}
\texttt{ground\_state} or even an \texttt{AdcMatrix},
which allows to share and re-use previously computed quantities,
such as the T2 amplitudes.
In lines 35 to 40, the obtained excited states are broadened
with a Lorentzian (width parameter $\gamma = 0.01$ atomic units) and plotted.
For this the \texttt{plot\_spectrum} function of \adcc
integrates with \matplotlib placing the spectrum directly
on a \matplotlib figure.
In this way the plot can be extended via the usual \matplotlib functionality.
In this example, we add a legend with \texttt{plt.legend()}
and display the image shown on the right of Figure \ref{fig:SpectrumSolverCascade}
with \texttt{plt.show()}.
% The observed pattern in the obtained excited states energies
% is in agreement with the discussion in the literature~\cite{Dreuw2014},
% namely that ADC(1) and ADC(3) give rise to higher-energy
% excitations compared to ADC(2),
% whereas ADC(2)-x is red-shifted.

The complete script with 40 lines including the code needed
for the visualisation of the excited states spectra is very tractable
and most lines of code are completely self-explanatory.
Obtaining key quantities such as the state densities as \numpy arrays
allows to easily extend the analysis and, e.g.,
visualise density differences in \matplotlib.
Such direct access to key quantities
greatly decreases the feedback loop between calculation and insight.
Furthermore, the brevity of the code
implies that it can be written spontaneously in an interactive
IPython~\cite{IPython} shell or a Jupyter notebook~\cite{Jupyter}
during a scientific discussion or a lecture.
This provides a powerful hands-on technique for
rapid ADC method development, debugging or interactive teaching sessions
directly in the web browser.

%%%%%%%%%%%%%%%%%%%%%
%--   Section B   --%
%%%%%%%%%%%%%%%%%%%%%
\section*{\sffamily \Large Flexible selection of frozen MOs and CVS spaces}
\label{sec:ex:cvs}

\begin{figure}[h]
	\centering
	\includegraphics[width=0.9\textwidth]{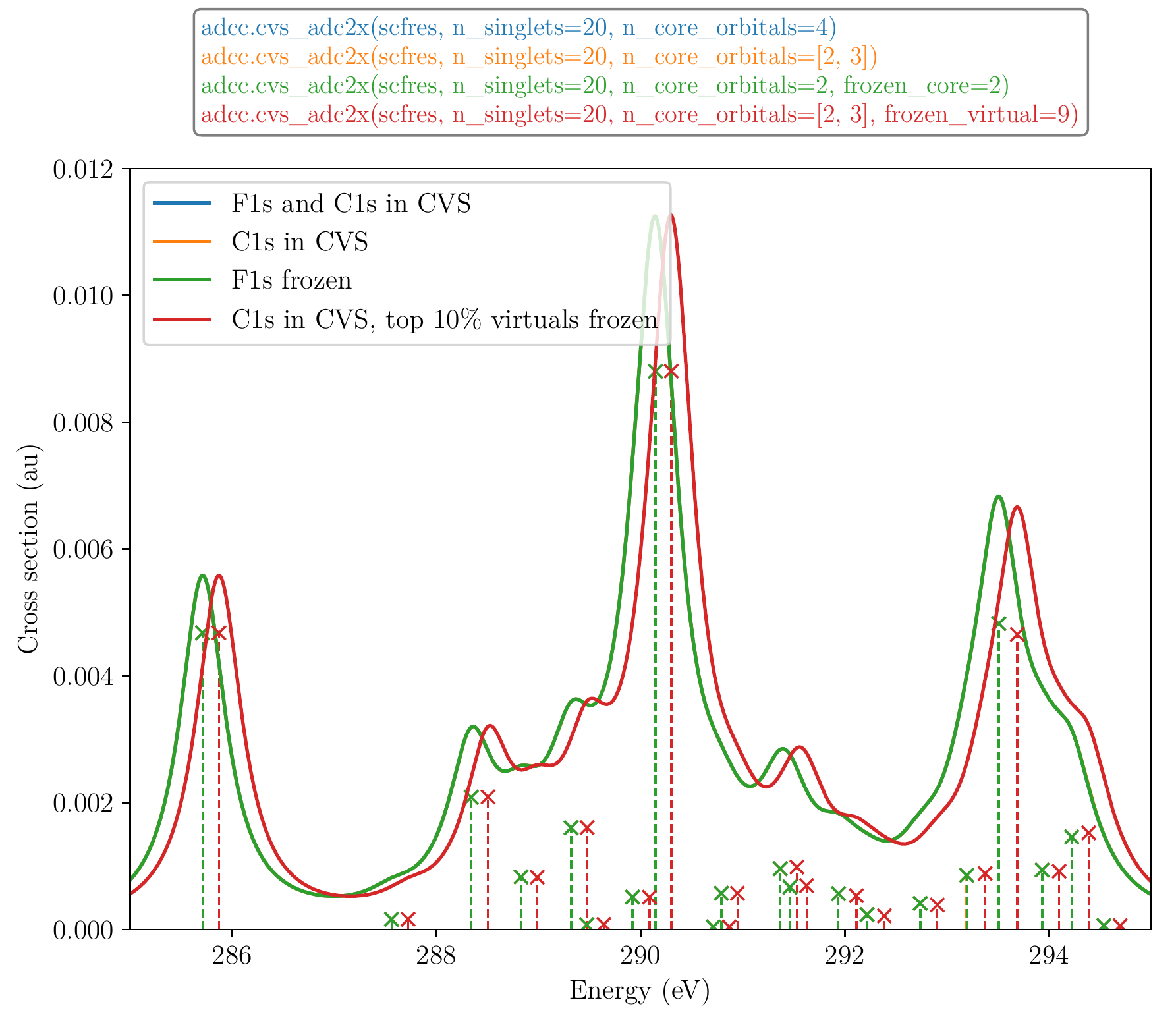}
	\caption{Carbon $K$-edge X-ray absorption spectrum of 1,1-difluoroethene,
	using different subspaces in CVS as well as by freezing core and/or virtual orbitals.
	Commands for calculating these spectra are shown above the Figure.}
	\label{fig:cvs_example}
\end{figure}

For the calculation of core-excited state, some flavour of the core-valence separation (CVS) approach is routinely employed to avoid the difficulty of considering states buried in a continuum of valence transitions~\cite{cvsadc1985, Trofimov2000, Wenzel2014a, cvseomccsd2019, cumulantXray2019}. Previous ADC implementation have, however, used a construction of the CVS space where all (core) MOs up to the last probed one is included in the CVS-ADC matrix. This leads to matrix dimensions which are larger than necessary, as we will now demonstrate using the \adcc implementation, which enables the use of any CVS spaces, as well as the freezing of arbitrary occupied or unoccupied MOs. We consider the carbon X-ray absorption spectrum of 1,1-difluoroethene, with results illustrated in Fig.~\ref{fig:cvs_example}. This system has been investigated experimentally~\cite{McLaren1987} and in theory~\cite{Fransson2013}, and possess significant shifts in transition energy due to the substitution of electronegative fluorine. The spectra have been calculated for a MP(2)/cc-pVTZ~\cite{Dunning1989} optimised structure,~\cite{Shao2015} with excited states calculated using CVS-ADC(2)-x/6-311++G**~\cite{6311Gstarstar}. This combination of ADC level and basis set has been noted to provide results in close agreement to experiment~\cite{Wenzel2014a,Wenzel2015}. Our results overestimates experimental measurement by $0.3-0.6$ eV~\cite{McLaren1987}, when scalar relativistic effects are accounted for.

We note that including only the two carbon core orbitals in the CVS space leads to identical results as including also the fluorines, at a lower computational cost. Alternatively, it is possible to freeze the fluorine core MOs and then select the lowest two MOs in the CVS space, which then again lead to identical results at a lower cost than including also the fluorines. The fluorine core MOs can be left outside the CVS space or frozen with identical results as they are spatially and energetically separated from the carbon core electrons, but this would not be the case if, e.g., the $L$-edge of heavier atoms are considered. Finally, we illustrate the use of a carbon-specific CVS space together with freezing the 10\% highest virtuals (here, 9 virtuals), which leads to an increase in transition energy of $0.16\pm0.01$ eV due to some lack of relaxation. Such a freezing of virtuals, as well as choosing CVS spaces focused on a single, chemically unique core orbital, can be employed to obtain additional lowering of computational cost. Care must be taken if any of these approaches are applied.

%%%%%%%%%%%%%%%%%%%%%
%--   Section C   --%
%%%%%%%%%%%%%%%%%%%%%
\section*{\sffamily \Large Solvent shift of nile red}
\label{sec:ex:nile_red}

\begin{figure}[h]
	\centering
	\includegraphics[width=1\textwidth]{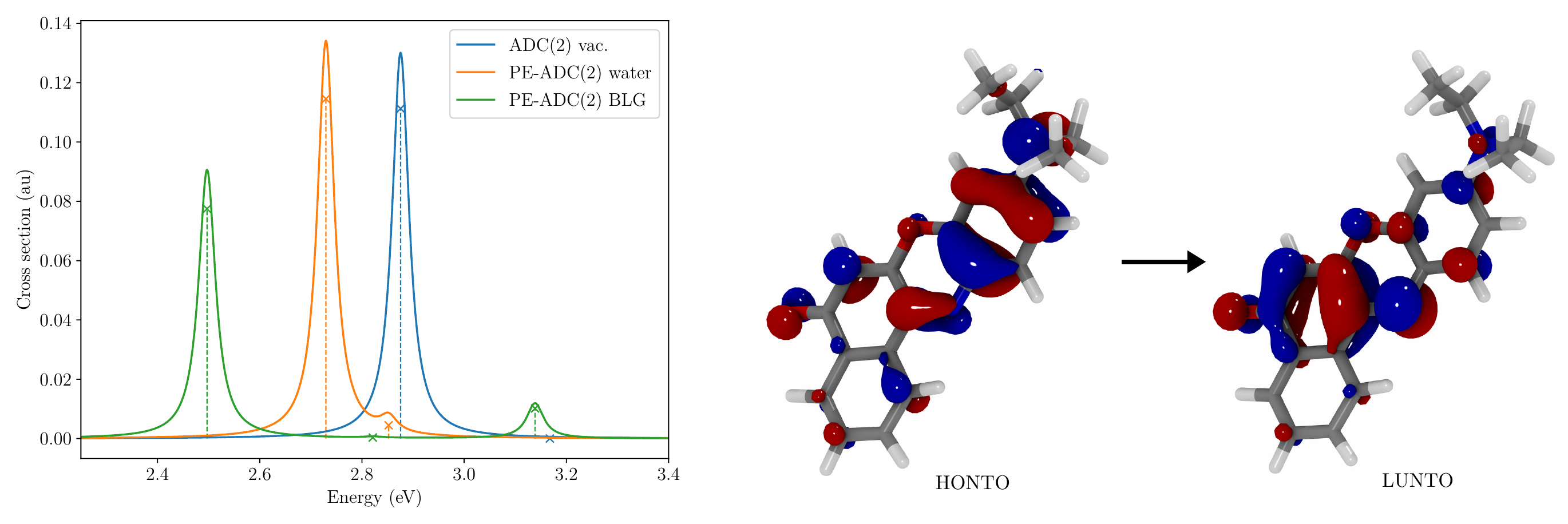}
	\caption{
		Absorption spectrum (left) with three singlet excited states of nile red
		in vacuum, water, and BLG. Highest occupied NTO (HONTO) and
		lowest unoccupied NTO (LUNTO) for the first singlet excited
		state in water (right).
		Spectra are broadened empirically with a Lorentzian ($\gamma = 0.02$ eV).
	}
\label{fig:nile_red}
\end{figure}

To illustrate the capabilities of \adcc in a biomolecular
application, the solvent shift of nile red
in water and protein environment is modelled using the polarisable
embedding (PE) model in combination with ADC (PE-ADC) \cite{Scheurer2018}.
Geometries of the chromophore and parameters
for the water and the \textbeta-lactoglobulin (BLG) protein environments
were used from previous work \cite{Scheurer2019}.
Using \pyscf as host program, three singlet excited states were computed
at the ADC(2)/cc-pVDZ \cite{Dunning1989} level of theory using \adcc.
Calculations included 420 basis functions and ran for 8 hours on 32 cores on a single node
(see supporting information for hardware details).
The PE-HF ground state calculations in \pyscf employed \texttt{cppe} \cite{Scheurer2019},
which is also a modular \python library.
Perturbative corrections of the excitation energies~\cite{Scheurer2018} were computed
directly in the \python job scripts.
The resulting spectra for nile red in vacuum, water, and BLG were obtained by convolution
with a Lorentzian function ($\gamma = 0.02$ eV).
Natural transition orbitals (NTOs) were also generated in the \python job script
by decomposing the AO transition density matrix with \numpy (\texttt{numpy.linalg.svd}) and
subsequently writing the orbitals to disk with the \texttt{cubegen} utility in \pyscf.
Finally, cube files of NTOs were rendered with VMD \cite{Humphrey1996}.

The resulting excitation spectra and NTOs of the
first excited states of nile red embedded in water
are depicted in Figure \ref{fig:nile_red}.
The NTOs clearly show the $\pi\pi^*$ nature of the lowest transition.
Consequently, the absorption cross section is high, and the transition
is strongly red-shifted in the embedded systems compared to vacuum.
Rather than discussing the properties of nile red, which has been done in
previous work \cite{Scheurer2019},
the given example shows that \adcc is capable of tackling systems of medium size
with good performance. As explained before, full flexibility is still granted,
here by a) computing perturbative corrections to the ADC excitation energies and b)
generating NTOs in user code with negligible programming effort.
In the same manner, other solvent models available through the supported host programs could
be combined with \adcc as well, e.g., continuum solvation models.
Furthermore, users could expand their scripts with more advanced
wave function analysis \cite{Plasser2014}, making the analysis more interactive
and tailored to the problem at hand.

\section*{\sffamily \Large CURRENT STATE AND SUPPORTED FEATURES}
\label{sec:state}
Presently the \adcc code base allows to model
electronically excited states using
various levels of the algebraic-diagrammatic construction scheme
for the polarisation propagator~(ADC).
This includes ADC(2), ADC(2)-x, and ADC(3)~\cite{Schirmer1982,Trofimov1999}
for the treatment of valence-excited states
as well as the respective core-valence separation~(CVS) variants%
~\cite{Trofimov2000,Wenzel2014b} for tackling core-excited states.
Both restricted as well as unrestricted Hartree-Fock references are supported
and few-reference ground states as well as their excitations can be approached
using the spin-flip modification~\cite{Lefrancois2015} of ADC.
For reducing the number of active occupied or virtual orbitals
and thus lowering the computational cost,
the frozen-core and frozen-virtual~\cite{Yang2017} approximations
can be additionally applied to all methods implemented in \adcc.
The selection of occupied or virtual orbitals to be frozen
as well as the selection of the core space is completely arbitrary,
i.e., not limited to contiguous blocks of occupied or virtual orbitals.
The tensor operations required in the ADC working equations
are evaluated in the \cpp core of \adcc utilising the
block-sparse tensor library \libten~\cite{Libtensor}
for exploiting symmetry and parallelising operations.
As a result \adcc is able to easily address medium-sized problems,
such as the cc-pVDZ ADC(2) calculations of nile red shown previously.

In \adcc ADC calculations are started and controlled from a \python module,
which exposes predefined ADC workflows for all aforementioned methods.
In line with the building-block approach taken by \adcc,
the Hartree-Fock~(HF) reference needs to be prepared in an external host program.
Currently four host programs are supported out-of-the box,
such that in practice the respective SCF datastructures are directly
understood by \adcc,
namely \psifour~\cite{Parrish2017,Smith2018}, \pyscf~\cite{Sun2017},
\molsturm~\cite{molsturmDesign} and
\veloxchem~\cite{Veloxchem}.
Further programs can be added by implementing about 20 functions
from \python, by employing a dictionary, or by using an HDF5 file
to pass precomputed data from the host program to \adcc.

The default workflow of \adcc computes a set of requested excited states
and offers \python functions for simple post-processing,
such as plotting of the excitation spectrum.
Key individual ADC quantities, such as transition or state properties
or respective transition or state density matrices
can be directly accessed as \numpy arrays~\cite{Walt2011}.
Beyond this, any individual step of \adcc and its intermediate results
may be requested from \python.
This allows unusual or novel ADC computations to be easily realised
and simplifies the extension of \adcc beyond its present capabilities in user code.
%This has been shown in two examples.
%Firstly, Section \textit{Comparison of ADC methods}
%presented a \python script of about 40 lines of code,
%which performed calculations using multiple ADC levels on the same
%molecule, each time utilising the cruder level of theory as a guess.
%This script, albeit short, provided direct feedback
%in the form of visualising the differing excitation spectra.
For a laptop-scale problem,
this makes \adcc highly suitable for interactive use,
allowing to grasp the quantum-chemical properties of a system
through one's own code rather than
through a ``black box''
provided by traditional program packages.
This greatly facilitates a hands-on approach to computational spectroscopy
for teaching or research.

%Secondly, Section \textit{Solvent Shift of Nile Red}
%demonstrated the interplay of three different
%\python-based packages in order to model the solvent shift of nile red
%in water and protein environment.
%For this the polarisable embedding model from CPPE~\cite{Scheurer2019}
%was used on top of a solvent-relaxed HF reference from \pyscf~\cite{Sun2017}.

%Not only the computational workflow, but also the iterative
%solver algorithms are implemented in \python.
%Right now the Jacobi-Davidson~\cite{Davidson1975}
%and the power method~\cite{Saad2011} are available.
%The latter being only a proof of principle,
%showing how novel numerical approaches for ADC
%can be rapidly prototyped inside the \adcc code in the future.

\section*{\sffamily \Large SUMMARY AND OUTLOOK}
The hybrid \cpp / \python module \adcc
for the simulation of excited states based on the algebraic-diagrammatic
construction for the polarisation propagator (ADC) has been presented.
Instead of aiming for a complete framework for spectroscopy simulations,
our philosophy is to integrate with existing software as much as possible
and only provide a single building block, i.e.,
a module for ADC calculations.
For this reason our \cpp core layer mainly contains code to interface with two libraries,
\libten~\cite{Libtensor}
for performing tensor operations
and \libadc~\cite{Wormit2014} for the implementation of the ADC working equations.
Similarly, on the \python layer we aim to integrate
both with the conventional \python ecosystem, i.e.,
with libraries such as \numpy~\cite{Walt2011} or \matplotlib~\cite{Matplotlib},
as well as \python-based quantum chemistry software:
Running calculations with \adcc by supplying
a Hartree-Fock reference from \pyscf~\cite{Sun2017}, \psifour~\cite{Parrish2017,Smith2018},
\molsturm~\cite{molsturmDesign}, or
\veloxchem~\cite{Veloxchem} is supported out of the box.
Adcc is also one integral part of the Gator program
for spectroscopy simulations using correlated wavefunctions~\cite{Gator}.

The required interfaces, both on the \cpp and the \python layer,
are kept simple and are well-documented
with ideally all functionality of the \cpp core being
available from \python as well.
This has the advantage that large parts of \adcc,
including our iterative solver algorithms,
could be implemented in \python.
The result is a flexible module with extensible workflows,
which has been demonstrated in the given examples.
Despite this flexibility,
\adcc easily performs calculations with about 400 basis functions
making \adcc not only a useful tool for
method development, but also for practical research calculations or teaching.

In the future we plan to extend \adcc to other ADC methods
of similar mathematical structure,
such as \mbox{IP-ADC}~\cite{Schirmer1983,vonNiessen1984,Schirmer1989,Dempwolff2019}.
To reduce the current memory requirements we want to resort
to factorisation techniques for the electron-repulsion integral tensor%
~\cite{Beebe1977,Pedersen2009,Aquilante2011}
or extend our tensor interface to simplify batched operations~\cite{Libtensor}
and disk-based algorithms.
For this we expect our focus on \python and open interfaces
to accelerate developments.
A clear aim is also to enhance deeper integration of \adcc into
other quantum-chemical software projects.
This has recently been achieved for \psifour~\cite{Parrish2017},
where ADC methods from \adcc can be directly used in \psifour inputs.
As a result not only the feature set of each involved project would grow,
but the implied possibility to mix and match
software building blocks for a scientific simulation at wish,
generates an environment for sustainable scientific innovation.

% Add here reference to Veloxchem and Gator for further work.

\section*{\sffamily \Large FUNDING INFORMATION}
MS was supported by the Deutsche Forschungsgemeinschaft (DFG) by means of
the research training group ``CLiC" (GRK 1986, Complex Light Control).
TF was supported by a grant from the Swedish Research Council (Grant No.\ 2017-00356).
This work was supported by the
Heidelberg Graduate School of Mathematical and Computational
Methods for the Sciences (GSC220).

\section*{\sffamily \Large RESEARCH RESOURCES}
Calculations in this work were supported by the state of Baden-Württemberg through bwHPC
(bwForCluster MLS\&WISO) and the German Research Foundation (DFG) through grant INST 35/1134-1 FUGG.

\section*{\sffamily \Large ACKNOWLEDGMENTS}
The authors thank Patrick Norman
for stimulating the redesign of \adcc from an earlier version of the code
and Adrian L. Dempwolff for discussions and helpful comments during the preparation
of the manuscript.

\section*{\sffamily \Large FURTHER READING}
The details of the \python interface of \adcc
as well as guides for installation and for getting started
can be found in the \adcc documentation.
The \adcc documentation is available online
at \url{https://adc-connect.org}.

% end parts

\bibliography{literature.bib}
\end{document}